\documentclass[a4paper,english,british,superscriptaddress,prl,twocolumn]{revtex4-1}
\usepackage{mathpazo}
\usepackage[scaled=0.86]{berasans}
\usepackage[T1]{fontenc}
\usepackage[utf8]{inputenc}
\usepackage{color}
\usepackage{babel}
\usepackage{amsthm}
\usepackage{amsmath}
\usepackage{amssymb}
\usepackage{graphicx}
\usepackage{esint}
\usepackage[unicode=true,pdfusetitle,
 bookmarks=false,
 breaklinks=false,pdfborder={0 0 0},backref=section,colorlinks=true]
 {hyperref}
\hypersetup{
 citecolor=blue,urlcolor=blue}
\usepackage{breakurl}

\makeatletter

\@ifundefined{textcolor}{}
{%
 \definecolor{BLACK}{gray}{0}
 \definecolor{WHITE}{gray}{1}
 \definecolor{RED}{rgb}{1,0,0}
 \definecolor{GREEN}{rgb}{0,1,0}
 \definecolor{BLUE}{rgb}{0,0,1}
 \definecolor{CYAN}{cmyk}{1,0,0,0}
 \definecolor{MAGENTA}{cmyk}{0,1,0,0}
 \definecolor{YELLOW}{cmyk}{0,0,1,0}
}
 \theoremstyle{definition}
 \newtheorem*{defn*}{\protect\definitionname}
  \theoremstyle{remark}
  \newtheorem*{rem*}{\protect\remarkname}

\usepackage{babel}
\usepackage{amsthm}\usepackage{bm}\usepackage{amsfonts}\usepackage{mathrsfs}\usepackage{bbm}%Usefull math packages
\usepackage{xspace}%Useful to add space in macros
\usepackage{pgfplots}%Useful to plot graphs with latex

\def\be{\begin{equation}}
\def\ee{\end{equation}}
\def\bea{\begin{eqnarray}}
\def\eea{\end{eqnarray}}
\def\ben{\begin{equation*}}
\def\een{\end{equation*}}
\def\bean{\begin{eqnarray*}}
\def\eean{\end{eqnarray*}}
\def\bma{\begin{mathletters}}
\def\ema{\end{mathletters}}
\def\bi{\begin{itemize}}
\def\ei{\end{itemize}}

\makeatother

  \addto\captionsbritish{\renewcommand{\definitionname}{Definition}}
  \addto\captionsbritish{\renewcommand{\remarkname}{Remark}}
  \addto\captionsenglish{\renewcommand{\definitionname}{Definition}}
  \addto\captionsenglish{\renewcommand{\remarkname}{Remark}}
  \providecommand{\definitionname}{Definition}
  \providecommand{\remarkname}{Remark}

\begin{document}

\title{Reciprocal ontological models show indeterminism of the order of
quantum theory}

\author{Somshubhro Bandyopadhyay}

\email{som@jcbose.ac.in}

\affiliation{Department of Physics and Center for Astroparticle Physics and Space
Science, Bose Institute, Block EN, Sector V, Bidhan Nagar, Kolkata
700091, India}

\author{Manik Banik}

\email{manik11ju@gmail.com}

\affiliation{Physics and Applied Mathematics Unit, Indian Statistical Institute,
203 B. T. Road, Kolkata 700108, India}

\author{Some Sankar Bhattacharya}

\email{somesankar@gmail.com}

\affiliation{Physics and Applied Mathematics Unit, Indian Statistical Institute,
203 B. T. Road, Kolkata 700108, India}

\author{Sibasish Ghosh}

\email{sibasish@imsc.res.in}

\affiliation{Optics \& Quantum Information Group, The Institute of Mathematical
Sciences, C.I.T Campus, Tharamani, Chennai 600 113, India}

\author{Guruprasad Kar}

\email{gkar@isical.res.in}

\affiliation{Physics and Applied Mathematics Unit, Indian Statistical Institute,
203 B. T. Road, Kolkata 700108, India}

\author{Amit Mukherjee}

\email{amitisiphys@gmail.com}

\affiliation{Physics and Applied Mathematics Unit, Indian Statistical Institute,
203 B. T. Road, Kolkata 700108, India}

\author{Arup Roy}

\email{arup145.roy@gmail.com}

\affiliation{Physics and Applied Mathematics Unit, Indian Statistical Institute,
203 B. T. Road, Kolkata 700108, India}
\begin{abstract}
The question whether indeterminism in quantum measurement outcomes
is fundamental or is there a possibility of constructing a finer theory
underlying quantum mechanics that allows no such indeterminism, has
been debated for a long time. We show that within the class of ontological
models due to Harrigan and Spekkens, those satisfying \emph{preparation-measurement
reciprocity} must allow indeterminism of the order of quantum theory.
Our result implies that one can design quantum random number generator,
for which it is impossible, even in principle, to construct a reciprocal
deterministic model. 
\end{abstract}
\maketitle

\section{Introduction}

Quantum mechanics is believed to be fundamentally random. According
to quantum theory, measurement outcomes on a quantum system prepared
in a known state cannot be definitively predicted as long as the state
is not an eigenstate of the observable being measured \cite{Born'1926,Neumann'1955}.
This is said to be quantum indeterminism, which gives rise to quantum
randomness \cite{Jennewein'2000,Stefanov'2000,Atsushi'2008}. The
significance of quantum randomness lies in the fact that true randomness
is hard to characterize mathematically \cite{Chaitin'1977,Knuth'1981},
and also cannot be obtained from classical physics \cite{Butterfield'1998}.
Thus quantum randomness, regarded as the only form of true randomness
in nature, becomes crucial as a resource for applications ranging
from cryptography to numerical simulation of physical and biological
systems. 

On the other hand, we know that classical statistical physics also
allows indeterminism, and therefore, randomness of some form. This
randomness, however, cannot be considered genuine because the underlying
theory, i.e., Newtonian physics is deterministic. In fact, the source
of randomness in a purely classical theory can be attributed to our
lack of knowledge. This observation alone makes room for a similar
argument in the case of quantum theory. That is, if there is a finer
\emph{deterministic} theory that underlies quantum mechanics, then
quantum randomness ceases to be fundamental. Indeed, for two level
quantum systems such theories do exist, e.g., Bell-Mermin model \cite{Bell'1966,Mermin'1993},
and Kochen-Specker model \cite{KS'1967}. Therefore, the question
of existence of a finer deterministic theory, which nonetheless should
reproduce quantum statistics of measurement outcomes, becomes important. 

The above question can be addressed within the framework of ontological
models due to Harrigan and Spekkens \cite{Harrigan'2010}. It is well
known that one classification of ontological models arise from the
probable interpretations of quantum state $\vert\psi\rangle$; i.e.,
whether $\left|\psi\right\rangle $ represents the physical reality
or merely the observer's knowledge about the quantum system. In fact,
the question of interpretation of quantum state has been strongly
debated since the inception of quantum theory \cite{Ballentine'1970,Bell'1966,Bohm'1952,Bohr'1935,EPR'1935,Popper'1967,Peres'1984}. 

An ontological model is said to be $\psi$\emph{-epistemic} if it
considers $\vert\psi\rangle$ to represent observer's knowledge about
the system, and $\psi$\emph{-ontic} if it considers $\left|\psi\right\rangle $
to represent reality of the system, e.g., \cite{Saunders'2010,Bohm'1993,Holland'1993,Bassi'2003}.
Thus, in a $\psi$\emph{-epistemic} model we can find distinct quantum
states with overlapping probability distributions in the space of
ontic states, whereas in a $\psi$\emph{-ontic} model distinct quantum
states correspond to probability distributions that do not overlap.
Moreover, we say that a model is $maximally\;\psi$\emph{-epistemic}
if the quantum overlap between two state vectors can be completely
accounted for by the overlap of their probability distributions in
the ontic space. 

The above ontological models can be characterized by the so called
\textquotedbl{}degree of epistemicity\textquotedbl{} $0\leq\Omega\left(\psi,\phi\right)\leq1$
defined for a pair of quantum states $\left(\left|\psi\right\rangle ,\left|\phi\right\rangle \right)$
\cite{Maroney'2012(1),Maroney'2012(2)}. In particular, we have $\Omega\left(\psi,\phi\right)=0$
for a $\psi$-ontic model, $\Omega\left(\psi,\phi\right)=1$, for
a maximally $\psi$-epistemic model and $0<\Omega\left(\psi,\phi\right)<1$
for models that are neither ontic nor maximally $\psi$-epistemic,
also called non-maximally $\psi$-epistemic \cite{Aaronson'2013,Lewis'2012}. 

The motivation of the present work stems from a recent result by Maroney
\cite{Maroney'2012(1)}. He proved a powerful theorem, which states
that for quantum systems of dimension greater than two it is impossible
to construct a maximally $\psi-$epistemic ontological model. Maroney's
result stands out because it was proved without auxiliary assumptions
unlike results \cite{PBR'2012,Colbeck'2012} claiming $\vert\psi\rangle$
must be ontic. For example, the authors of \cite{Colbeck'2012} have
shown that a quantum system's wave function is in one-to-one correspondence
with its elements of reality, which implies that quantum indeterminism
is \emph{irreducible}; however, this was derived under a strong\emph{
freedom of choice} assumption whose characterization is unrealistic
and unsatisfactory \cite{Ghirardi'2013,Ballentine'2014}. Also note
that \cite{PBR'2012} requires additional property, namely, preparation-independence,
and without it explicit counter-examples show that epistemic models
underlying quantum theory can be formulated \cite{Lewis'2012,Aaronson'2013}. 

To understand the implication of Maroney's theorem, we first need
to briefly describe the structural features associated with ontological
models. These features were explicitly discussed by Ballentine \cite{Ballentine'2014}.
He introduced the property \emph{preparation-measurement reciprocity},
which is satisfied by quantum theory but not necessarily holds in
an ontological model. It's given by two conditions: The first one,
termed as, \textquotedbl{}Quantum certainty\textquotedbl{} states
that a quantum state $\left|\psi\right\rangle $ will always pass
the measurement filter $\left|\psi\left\rangle \right\langle \psi\right|$
(here we note that any ontological model that aims to reproduce quantum
statistics must satisfy quantum certainty). The second condition is
simply the converse, i.e., $\left|\psi\right\rangle $ is the only
state that passes the filter $\left|\psi\left\rangle \right\langle \psi\right|$
with probability one. An ontological model that satisfies these two
conditions is said to be \emph{reciprocal}, and \emph{non-reciprocal}
if the converse does not hold (this is because we always require that
any ontological model should reproduce quantum statistics). Furthermore,
an ontological model is \emph{outcome-deterministic} if the measurement
outcomes, at the ontological level, can be predicted with certainty,
else the model is said to be \emph{outcome-indeterministic}. 

Ballentine \cite{Ballentine'2014} showed that Maroney's theorem rules
out ontological models that are both reciprocal and outcome-deterministic.
In other words, a reciprocal ontological model must be indeterministic.
Naturally, the question is, To what extent such a model is indeterministic?
In this paper, we focus on this question. 

We prove that indeterminism in a reciprocal ontological model is of
the order of quantum theory. This is shown in two steps. For a quantum
state $\left|\psi\right\rangle $ and a projective observable $\vert\phi\rangle\langle\phi\vert$,
let $\mathcal{I}_{{\rm ont}}\left(\psi,\phi\right)$ and $\mathcal{I}_{{\rm Q}}\left(\psi,\phi\right)$
(precisely defined later) quantify indeterminism associated with outcome
$\phi$ in a reciprocal ontological model and quantum theory respectively.
We first show that the following relation holds: 
\begin{eqnarray}
\mathcal{I}_{{\rm ont}}\left(\psi,\phi\right) & = & \left[1-\Omega\left(\psi,\phi\right)\right]\mathcal{I}_{{\rm Q}}\left(\psi,\phi\right),\label{main-result}
\end{eqnarray}
where $\Omega\left(\psi,\phi\right)$ is the degree of epistemicity
of the pair of quantum states $\left(\left|\psi\right\rangle ,\left|\phi\right\rangle \right)$.
Note that, although Eq.$\,$(\ref{main-result}) connects $\mathcal{I}_{{\rm ont}}$
and $\mathcal{I}_{{\rm Q}}$ via the degree of epistemicity of the
states, it does not tell us how $\mathcal{I}_{{\rm ont}}$ compares
with $\mathcal{I}_{{\rm Q}}$. To compare, we therefore need to know
$\Omega\left(\psi,\phi\right)$ for a given pair of states $\left(\left|\psi\right\rangle ,\left|\phi\right\rangle \right)$. 

In the next step, we make use of a basis independent upper bound on
the degree of epistemicity \cite{Maroney'2012(1)} to show that for
all pairs of non-orthogonal quantum states in Hilbert spaces of dimension
$d\geq3$, 
\begin{eqnarray}
\mathcal{I}_{{\rm ont}} & \geq & \frac{\left(d-2\right)^{2}}{d^{2}+\left(d-2\right)^{2}}\mathcal{I}_{{\rm Q}}.\label{maininequality}
\end{eqnarray}
For example, in three dimension, $\mathcal{I}_{{\rm ont}}\geq\frac{1}{10}\mathcal{I}_{{\rm Q}}$,
and in the limit of large dimension of the Hilbert space, $\mathcal{I}_{{\rm ont}}\left(\psi,\phi\right)\geq\frac{1}{2}\mathcal{I}_{{\rm Q}}\left(\psi,\phi\right).$
We therefore conclude that \emph{indeterminism in a reciprocal ontological
model is of the order of quantum theory}. Note that our result implies
that it is possible to design quantum random number generator for
which it is not possible, even in principle, to construct a deterministic
ontological model. 

We further observe that Eq.$\,$(\ref{main-result}) establishes the
following correspondence: 

$\mathcal{I}_{{\rm ont}}\left(\psi,\phi\right)=\mathcal{I}_{{\rm Q}}\left(\psi,\phi\right)$\emph{$\iff$$\psi$-ontic}; 

$\mathcal{I}_{{\rm ont}}\left(\psi,\phi\right)=0$\emph{$\iff$maximally
$\psi$-epistemic}; 

$\mathcal{I}_{{\rm ont}}\left(\psi,\phi\right)<\mathcal{I}_{{\rm Q}}\left(\psi,\phi\right)$\emph{$\iff$nonmaximally
$\psi$-epistemic}. \\

The paper is arranged in the following way. Sections II and III contain
the necessary background material. In Sec.$\,$II we first define
the notion of indeterminism in an operational theory. We then describe,
in some detail, quantum indeterminism and the condition of preparation-measurement
reciprocity. In Sec.$\,$III we first briefly review the general theory
of ontological models, and the ontological models of quantum theory
following Ballentine \cite{Ballentine'2014}. We also discuss Maroney's
no-go theorem \cite{Maroney'2012(1)}, and its implication as pointed
out by Ballentine \cite{Ballentine'2014}. In Sec.$\,$IV we define
the notion of indeterminism in an ontological model of operational
quantum theory, and prove our results given by Eqs.$\,$(\ref{main-result})
and (\ref{maininequality}). Sec.$\,$V concludes with a discussion
on related aspects.

\section{Indeterminism in operational quantum theory }

The primitive elements of an operational theory are preparation procedures
$P\in\mathcal{P}$, transformations $T\in\mathcal{T}$, and measurement
procedures $M\in\mathcal{M}$, where $\mathcal{P},\mathcal{T}$ and
$\mathcal{M}$ denote collection of \emph{all} permissible preparations,
transformations and measurements respectively. An operational theory
specifies the probabilities of different outcomes of a measurement
performed on a system prepared according to some definite procedure.
Let $p(k|P,M)\in\left[0,1\right]$ denote the probability of outcome
$k$ when a measurement $M\in\mathcal{M}$ is performed on a system
prepared according to some procedure $P\in\mathcal{P}$ \cite{Spekkens'2005}.
We therefore have, 
\begin{eqnarray}
\sum_{k\in\mathcal{K}_{M}}p\left(k\vert P,M\right) & = & 1\;\forall P,M,\label{sec2-eq-1}
\end{eqnarray}
where $\mathcal{K}_{M}$ denotes the set of all measurement outcomes
corresponding to the measurement $M$. 
\begin{defn*}
An operational theory is \emph{deterministic} if and only if $p(k|P,M)$
takes values either $0$ or $1$ for every $k,P,M$. Otherwise the
theory is \emph{indeterministic}. 
\end{defn*}
In operational quantum theory, a preparation corresponds to the state
of a quantum system described by a vector $|\psi\rangle\in\mathcal{H}$,
where $\mathcal{H}$ is the Hilbert space associated with the system;
a measurement corresponds to an hermitian operator -- an observable
$\mathcal{O}$ -- acting on $\mathcal{H}$, whose eigenvalues represent
the measurement outcomes. Suppose $\left\{ \phi_{k}\right\} $ is
the set of eigenvalues of $\mathcal{O}$ with the corresponding set
of eigenvectors $\left\{ \vert\phi_{k}\rangle\right\} $. When observable
$\mathcal{O}$ is measured on the system prepared in state $|\psi\rangle$,
the probability of obtaining the outcome $\phi_{k}$ is given by the
Born rule: 
\begin{eqnarray}
p\left(\phi_{k}\vert\psi,\mathcal{O}\right) & = & \left|\left\langle \phi_{k}\vert\psi\right\rangle \right|^{2},\label{equantum-prob}
\end{eqnarray}
where $p\left(\phi_{k}\vert\psi,\mathcal{O}\right)\in\left[0,1\right]$.
For a specific outcome, the probability is one if and only if $\vert\psi\rangle$
is the corresponding eigenstate and is zero if and only if $\vert\psi\rangle$
is orthogonal to the corresponding eigenstate. It therefore follows
that quantum theory in general is indeterministic. 

To define indeterminism in quantum theory we proceed as follows. We
first observe that a measurement (observable $\mathcal{O}$) is, in
fact, a collection of projective measurements (observables) $\left\{ \vert\phi_{k}\rangle\langle\phi_{k}\vert\right\} $.
Therefore, without loss of generality, we can represent a measurement
by a single projective observable: $M_{\phi}=\vert\phi\rangle\langle\phi\vert$.
Stated in this language, the indeterminism in quantum mechanics for
a preparation-measurement pair $\left(\left|\psi\right\rangle ,M_{\phi}\right)$
is defined as, 
\begin{equation}
\mathcal{I}_{{\rm Q}}(\psi,\phi):=|\langle\psi|\phi\rangle|^{2},\label{quantum-indterminism}
\end{equation}
where $\mathcal{I}_{{\rm Q}}\in[0,1]$, and the outcome is indeterministic
if and only if $\mathcal{I}_{{\rm Q}}\ne0,1$. 

It is now easy to see that when measurement of an observable $\mathcal{O}$
on a system in state $|\psi\rangle$ results in indeterministic outcomes,
the preparation-measurement pair $\left(\psi,\mathcal{O}\right)$
can be used to design quantum random number generator, where randomness
can be quantified by the\emph{ guessing probability} $\mathcal{G}:=\max_{\phi_{k}}p\left(\phi_{k}\vert\psi,\mathcal{O}\right)$
\cite{Acin'2012}. Note that the amount of randomness can also be
quantified in terms of bit, i.e., $H_{\infty}=-\log_{2}\mathcal{G}$,
which is the expression for min entropy \cite{Koenig'2009}.

\subsection*{Preparation-measurement reciprocity }

A consequence of (\ref{quantum-indterminism}) is an interesting reciprocal
relationship of state preparation and measurement as observed by Ballentine
\cite{Ballentine'2014}: 
\begin{itemize}
\item \emph{Quantum Certainty:} A system that is prepared in the state $\vert\psi\rangle$
will always pass the test of measuring the projector $\vert\psi\rangle\langle\psi\vert$. 
\item \emph{Converse:} The state $\vert\psi\rangle$ is the only state that
will pass the projective measurement filter $\vert\psi\rangle\langle\psi\vert$
with certainty. 
\end{itemize}
Thus quantum mechanics admits \emph{preparation-measurement reciprocity},
or \emph{reciprocity} for short. It is important to recognize that
a finer theory, which may or may not be deterministic, must satisfy
\textquotedbl{}Quantum Certainty\textquotedbl{} otherwise it will
fail to reproduce quantum statistics of measurement outcomes. However,
it is possible that in such a theory the converse may not hold.

\section{Ontological models and degree of epistemicity }

In an ontological model of an operational theory, the primitives of
description are the properties of microscopic systems \cite{Harrigan'2010},
where a preparation procedure prepares a system with definite properties
and a measurement procedure tells us something about those properties.
A complete specification of the properties of a system is referred
to as the \emph{ontic state} --\emph{ the state of reality} of that
system and is denoted by $\lambda\in\Lambda$, where $\Lambda$ denotes
the space of ontic states. A particular preparation $P$ actually
yields an ontic state $\lambda$, and in general the same preparation,
when repeated, produces a different ontic state. We therefore have
a probability distribution $0\leq\mu(\lambda|P)\leq1$ over the ontic
states $\lambda\in\Lambda$, said to be the \emph{epistemic state
-- the state of knowledge,} which satisfies:
\begin{eqnarray}
\int_{\Lambda}\mu\left(\lambda\vert P\right) & = & 1\;\forall P.\label{epistemic-state}
\end{eqnarray}
 For a measurement $M$, the probability of obtaining an outcome $k$
is determined by a \emph{response (or indicator) function} $0\leq\xi(k|\lambda,M)\leq1$
satisfying 
\begin{eqnarray}
\sum_{k}\xi\left(k\vert\lambda,M\right) & = & 1\;\forall k,\lambda,M.\label{eepistmeic-response}
\end{eqnarray}

\begin{defn*}
If the response function always takes values either $0$ or $1$ then
the ontological model is \emph{outcome-deterministic.} Otherwise it
is \emph{outcome-indeterministic}.
\end{defn*}
The ontological model, however, is required to reproduce the predictions
of the operational theory for every preparation-measurement pair $\left(P,M\right)$.
We therefore must have, 
\begin{eqnarray}
p\left(k\vert P,M\right) & = & \int_{\Lambda}\xi\left(k\vert\lambda,M\right)\mu\left(\lambda\vert P\right)d\lambda\label{probability-muandresponse}
\end{eqnarray}

\begin{rem*}
If the ontological model is outcome-deterministic then the operational
theory may or may not be deterministic. However, if the operational
theory is deterministic, then it necessarily implies that at the ontological
level, the theory is outcome-deterministic. 
\end{rem*}

\subsection*{Ontological models of operational quantum theory }

In an ontological model of operational quantum theory we have, the
epistemic state $0\leq\mu\left(\lambda\vert\psi\right)\leq1\;\forall\lambda$,
satisfying 
\begin{eqnarray}
\int_{\Lambda}\mu\left(\lambda\vert\psi\right)d\lambda & = & 1\;\forall\,\vert\psi\rangle,\label{onto-quantum-1}
\end{eqnarray}
and the response function $0\leq\xi\left(\phi_{k}\vert\lambda,\mathcal{O}\right)\leq1$,
satisfying
\begin{eqnarray}
\sum_{\phi_{k}}\xi\left(\phi_{k}\vert\lambda,\mathcal{O}\right) & = & 1\;\forall\lambda,\mathcal{O}\label{onto-quantum-2}
\end{eqnarray}
We require that the ontological model must reproduce the predictions
of operational quantum theory. Therefore, for any projective observable
$M_{\phi}=\vert\phi\rangle\langle\phi\vert$, the ontological model
must satisfy 
\begin{equation}
\int_{\Lambda}\xi\left(\phi\vert\lambda\right)\mu\left(\lambda\vert\psi\right)d\lambda=\left|\left\langle \psi\vert\phi\right\rangle \right|^{2},\label{eq6-1}
\end{equation}
where for simplicity the following notation: $\xi\left(\phi\vert\lambda,M_{\phi}\right)=\xi\left(\phi\vert\lambda\right)$
is adopted and will hold for the rest of the paper. In the ontic state
space $\Lambda$ we can identify the following subsets \cite{Ballentine'2014}:
\begin{eqnarray}
\Lambda_{\psi}: & = & \{\lambda\in\Lambda~|~\mu(\lambda|\psi)>0\},\label{eq7-1}\\
\mbox{Supp}\left[\xi\left(\psi\vert\lambda\right)\right]: & = & \{\lambda\in\Lambda~|~\xi(\psi|\lambda)>0\},\label{eq8-1}\\
\mbox{ Core}\left[\xi\left(\psi\vert\lambda\right)\right]: & = & \{\lambda\in\Lambda~|~\xi(\psi|\lambda)=1\}.\label{eq9-1}
\end{eqnarray}
Eq.$\,$(\ref{eq6-1}) implies that an ontological model underlying
quantum theory must satisfy \emph{Quantum Certainty}, i.e., for the
ontic states $\lambda$'s that appear with a non-zero probability
in the preparation of the state $\vert\psi\rangle$, the response
function $\xi(\psi|\lambda)=1~\forall~\lambda\in\Lambda_{\psi}$.
Consequently, 
\begin{equation}
\int_{\Lambda_{\psi}}\xi(\psi|\lambda)\mu(\lambda|\psi)d\lambda=1.\label{eq10-1}
\end{equation}
Therefore, among the subsets defined in Eqs.$\,$(\ref{eq7-1}-\ref{eq9-1})
the following set inclusion relation holds, 
\begin{equation}
\Lambda_{\psi}\subseteq\mbox{Core}[\xi(\psi|\lambda)]\subseteq\mbox{Supp}[\xi(\psi|\lambda)].\label{eq11-1}
\end{equation}
The above inclusion relations immediately lead to the following classification. 
\begin{defn*}
An ontological model for quantum theory that satisfies preparation-measurement
reciprocity (Quantum Certainty and its Converse) is said to be \emph{reciprocal}.
If the Converse does not hold the model is said to be \emph{non-reciprocal}. \end{defn*}
\begin{rem*}
For a reciprocal ontological model the relation
\begin{eqnarray}
\Lambda_{\psi} & = & \mbox{Core}[\xi(\psi|\lambda)]~\forall~\vert\psi\rangle\label{eq:reciprocal}
\end{eqnarray}
holds. Note that quantum theory also satisfies the \emph{preparation-measurement
reciprocity} relation. \end{rem*}
\begin{defn*}
An ontological model for quantum theory is said to be \emph{outcome-deterministic}
if and only if $\mbox{Core}[\xi(\psi|\lambda)]=\mbox{Supp}[\xi(\psi|\lambda)]~\forall~\left|\psi\right\rangle $,
otherwise it is said to be \emph{outcome-indeterministic}. \end{defn*}
\begin{rem*}
If an ontological model assigns indeterministic response function
to some outcomes, then at least for those outcomes quantum randomness
exists at ontological level.
\end{rem*}
Thus essentially we have four kinds of ontological models. An ontological
model is either (a) reciprocal and deterministic/indeterministic or
(b) non-reciprocal and deterministic/indeterministic. We, however,
require all models to satisfy the condition of \textquotedbl{}Quantum
Certainty\textquotedbl{}.

\subsection*{Maroney's no go theorem and its implication}

Recently Maroney introduced the concept of \emph{degree of epistemicity}
\cite{Maroney'2012(1),Maroney'2012(2)} for ontological models of
quantum theory. He showed that in any ontological model the following
relation holds: 
\begin{equation}
\int_{\Lambda_{\phi}}\mu(\lambda|\psi)d\lambda=\Omega\left(\psi,\phi\right)|\langle\psi|\phi\rangle|^{2},\label{degree-of-spistemicity}
\end{equation}
where $0\le\Omega\left(\psi,\phi\right)\le1$ is defined as the ``degree
of epistemicity'' associated with the states $\left|\psi\right\rangle $
and $\left|\phi\right\rangle $ . As noted earlier, the ontological
models can be characterized by the degree of epistemicity. 

Without making auxiliary assumptions, Maroney proved that for Hilbert
spaces of dimension greater than two it is not possible to construct
a maximally $\psi$-epistemic ontological model $\left[\Omega\left(\psi,\phi\right)=1\right]$
for quantum theory \cite{Maroney'2012(1)}. In a very recent work
\cite{Ballentine'2014} Ballentine showed that if the ontological
model is maximally $\psi$-epistemic, then for any quantum state $\left|\psi\right\rangle $
the following relation must hold: 
\begin{equation}
\Lambda_{\psi}=\mbox{Core}\left[\xi\left(\psi\vert\lambda\right)\right]=\mbox{Supp}\left[\xi\left(\psi\vert\lambda\right)\right]\label{Ballentine-relation}
\end{equation}
The above equation tells us that a maximally $\psi$-epistemic model
must be reciprocal and outcome-deterministic. He further showed that
the converse also holds; therefore, for all $\left|\psi\right\rangle $
\begin{eqnarray}
\mbox{Maximally}~\psi-\mbox{epistemic}\Longleftrightarrow\mbox{Reciprocal}\nonumber \\
~~~~~~~~~~~~\mbox{\;\;\;\;\ AND~Outcome-deterministic}.
\end{eqnarray}
It follows that any ontological model can fail to be maximally $\psi$-epistemic
in only one of the three possible ways; it can be outcome-indeterministic
or non-reciprocal or both. 

Maroney's result implies that a reciprocal ontological model must
necessarily be indeterministic. In the following section we show that
the indeterminism is of the order of quantum theory.

\section{Indeterminism in reciprocal ontological model of quantum theory }

We begin by defining the notion of indeterminism in an ontological
model of quantum theory. To denote the subsets Core and Support associated
with a quantum state $\left|\eta\right\rangle $ we adopt the following
notations: $\mathcal{C}_{\eta}=\mbox{Core}\left[\xi\left(\eta\vert\lambda\right)\right]$
and $\mathcal{S}_{\eta}=\mbox{Supp}\left[\xi\left(\eta\vert\lambda\right)\right]$. 

For a given state preparation $\vert\psi\rangle$ and a measurement
$M_{\phi}=\vert\phi\rangle\langle\phi\vert$, define the region $\Lambda_{r}:=\Lambda_{\psi}\cap\left(\mathcal{S}_{\phi}\backslash\mathcal{C}_{\phi}\right)$.
Indeterminism in an ontological model is therefore defined as
\begin{equation}
\mathcal{I}_{{\rm ont}}(\psi,\phi):=\int_{\lambda\in\Lambda_{r}}\xi(\phi|\lambda)\mu(\lambda|\psi)d\lambda.\label{indeterminism-ontological-1}
\end{equation}
Observe that if the model is outcome-deterministic, then $\mathcal{S}_{\phi}=\mathcal{C}_{\phi}$,
and therefore, $\mathcal{I}_{{\rm ont}}=0$, i.e., at the ontological
level indeterminism is absent. It is also important to recognize that
while Eq.(\ref{quantum-indterminism}) quantifies indeterminism at
an operational level, Eq.$\,$(\ref{indeterminism-ontological-1})
does so at the ontological level. 

We now consider only reciprocal ontological models for quantum systems
in dimensions greater than two. From Maroney's theorem (discussed
in the previous section) we know that a reciprocal ontological model
must be outcome-indeterministic; i.e., for every quantum state $\left|\eta\right\rangle $
there exists a non-trivial region $\mathcal{S}_{\eta}\backslash\mathcal{C}_{\eta}\ne\emptyset$
in the ontic state space, where $0<\xi(\eta|\lambda)<1$. Thus for
a preparation-measurement pair $\left(\psi,M_{\phi}\right)$, we have
$\Lambda_{r}:=\Lambda_{\psi}\cap\left(\mathcal{S}_{\phi}\backslash\mathcal{C}_{\phi}\right)\neq\emptyset$.
This means, whenever the system is prepared in state $|\psi\rangle$,
outcome $\phi$ is indeterministic at the ontological level only for
those $\lambda\in\Lambda_{r}$. Since every $\lambda\in\Lambda_{r}$
gives indeterministic response for the outcome $\phi$, 
\begin{equation}
\mathcal{I}_{{\rm ont}}(\psi,\phi):=\int_{\lambda\in\Lambda_{r}}\xi(\phi|\lambda)\mu(\lambda|\psi)d\lambda>0.\label{indeterminism-ontological}
\end{equation}
\begin{figure}[t!]
\centering \includegraphics[width=7cm,height=4cm]{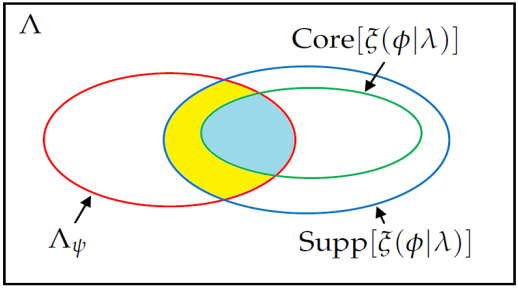} \protect\protect\caption{(Colour on-line) The blue shades region is $\Lambda_{\psi}\cap\mbox{Core}[\xi(\phi|\lambda)]$
and the yellow area is $\Lambda_{\psi}\cap\left(\mbox{Supp}[\xi(\phi|\lambda)]\setminus\mbox{Core}[\xi(\phi|\lambda)]\right)=\Lambda_{r}$.
Because we consider reciprocal models so $\Lambda_{\phi}=\mbox{Core}[\xi(\phi|\lambda)]$.}

\label{fig1} 
\end{figure}
We now prove Eq.$\,$(\ref{main-result}). We begin by recalling Eq.$\,$(\ref{eq6-1})
reproduced below, 
\begin{eqnarray}
\int_{\Lambda}\mu(\lambda|\psi)\xi(\phi|\lambda)d\lambda & = & |\langle\psi|\phi\rangle|^{2},\label{sec4-eq-2}
\end{eqnarray}
which simply captures the fact that an ontological model must reproduce
quantum statistics of measurement outcomes. Noting that the contribution
to the integral comes only from the region $\Lambda_{\psi}\cap\mathcal{S}_{\phi}$
(as it vanishes elsewhere), we can write Eq.$\,$(\ref{sec4-eq-2})
as 
\begin{eqnarray}
\int_{\Lambda_{\psi}\cap\mathcal{S}_{\phi}}\mu(\lambda|\psi)\xi(\phi|\lambda)d\lambda & = & |\langle\psi|\phi\rangle|^{2}.\label{sec4-eq-3}
\end{eqnarray}
We can now break up the region of the above integration in the following
way (see Fig.\ref{fig1}),
\begin{equation}
\int_{\Lambda_{\psi}\cap\mathcal{C}_{\phi}}\mu(\lambda|\psi)\xi(\phi|\lambda)d\lambda+\int_{\Lambda_{r}}\mu(\lambda|\psi)\xi(\phi|\lambda)d\lambda=|\langle\psi|\phi\rangle|^{2}.\label{sec4-eq-4}
\end{equation}
Using the fact that $\xi\left(\phi\vert\lambda\right)=1~\forall~\lambda\in\mathcal{C}_{\phi}$,
we have
\begin{equation}
\int_{\Lambda_{\psi}\cap\mathcal{C}_{\phi}}\mu(\lambda|\psi)d\lambda+\int_{\Lambda_{r}}\mu(\lambda|\psi)\xi(\phi|\lambda)d\lambda=|\langle\psi|\phi\rangle|^{2}.\label{sec4-eq-5}
\end{equation}
Because the concerned ontological model is reciprocal, i.e., $\mathcal{C}_{\phi}=\Lambda_{\phi}$,
\begin{equation}
\int_{\Lambda_{\psi}\cap\Lambda_{\phi}}\mu(\lambda|\psi)d\lambda+\int_{\Lambda_{r}}\mu(\lambda|\psi)\xi(\phi|\lambda)d\lambda=|\langle\psi|\phi\rangle|^{2}.\label{sec4-eq-6}
\end{equation}
Now observe that, $\int_{\Lambda_{\psi}\cap\Lambda_{\phi}}\mu(\lambda|\psi)d\lambda=\int_{\Lambda_{\phi}}\mu(\lambda|\psi)d\lambda$;
using this equality and Maroney's relation given by Eq.$\,$(\ref{degree-of-spistemicity}),
Eq.$\,$(\ref{sec4-eq-6}) can be expressed as,
\begin{eqnarray}
\int_{\Lambda_{r}}\mu(\lambda|\psi)\xi(\phi|\lambda)d\lambda & = & \left[1-\Omega\left(\psi,\phi\right)\right]\left|\left\langle \psi\vert\phi\right\rangle \right|^{2}.\label{sec4-eq-7}
\end{eqnarray}
Noting that the left hand side is simply the definition of $\mathcal{I}_{{\rm ont}}\left(\psi,\phi\right)$
given by Eq.$\,$(\ref{indeterminism-ontological}) and $\left|\left\langle \psi\vert\phi\right\rangle \right|^{2}=\mathcal{I}_{{\rm Q}}\left(\psi,\phi\right)$,
we finally arrive at our desired result, 
\begin{eqnarray}
\mathcal{I}_{{\rm ont}}\left(\psi,\phi\right) & = & \left[1-\Omega\left(\psi,\phi\right)\right]\mathcal{I}_{{\rm Q}}\left(\psi,\phi\right).\label{main-result-sec4}
\end{eqnarray}
The above equation is quite satisfying as it connects the notion of
indeterminism in quantum theory and reciprocal ontological model through
the degree of epistemicity. In order to compare $\mathcal{I}_{{\rm ont}}\left(\psi,\phi\right)$
with $\mathcal{I}_{{\rm Q}}\left(\psi,\phi\right)$ for a given pair
of states $\left(\left|\psi\right\rangle ,\left|\phi\right\rangle \right)$,
the knowledge of $\Omega\left(\psi,\phi\right)$ is required. We now
obtain inequality$\,$(\ref{maininequality}), which allows us to
compare $\mathcal{I}_{{\rm ont}}$ with $\mathcal{I}_{{\rm Q}}$. 

In \cite{Maroney'2012(1)}, Maroney noted that a basis independent
measure of degree of epistemicity allows one to set $\Omega\left(\psi,\phi\right)=\Omega\left(d\right)$,
where $\Omega\left(d\right)$ is a constant for all pairs of non-orthogonal
quantum states in dimensions $d\geq3$, and satisfies,
\begin{eqnarray}
\Omega\left(d\right) & \leq & \frac{d^{2}}{2d^{2}-4d+4}.\label{omega-upperbound}
\end{eqnarray}
Substituting the above in Eq.$\,$(\ref{main-result-sec4}) leads
us to 
\begin{eqnarray}
\mathcal{I}_{{\rm ont}} & \geq & \frac{\left(d-2\right)^{2}}{d^{2}+\left(d-2\right)^{2}}\mathcal{I}_{{\rm Q}}\label{main-inequality}
\end{eqnarray}
for all pairs of non-orthogonal states $\left(\left|\psi\right\rangle ,\left|\phi\right\rangle \right)$
in dimensions $d\geq3$. For instance, for quantum systems of dimension
three, we have $\mathcal{I}_{{\rm ont}}\geq\frac{1}{10}\mathcal{I}_{{\rm Q}}$,
and in the limit of large dimension $\mathcal{I}_{{\rm ont}}\geq\frac{1}{2}\mathcal{I}_{{\rm Q}}$.
It may be noted that if a basis independent measure of $\Omega\left(\psi,\phi\right)$
is not assumed, then one can find pairs of states $\left(\left|\psi\right\rangle ,\left|\phi\right\rangle \right)$
giving tighter upper bound on $\Omega\left(\psi,\phi\right)$, and
consequently a lower bound on $\mathcal{I}_{{\rm ont}}(\psi,\phi)$,
which is closer to $\mathcal{I}_{{\rm Q}}\left(\psi,\phi\right)$.
Thus we have shown that indeterminism in a reciprocal ontological
model must be of the order of quantum theory.

\section{Conclusions}

Quantum theory is known to be fundamentally random. This randomness
is due to the indeterminism associated with measurement outcomes when
an observable is being measured on a quantum system prepared in some
known state. Quantum randomness is significant as it is believed to
represent true randomness, and more so because true randomness does
not exist in classical physics. However, if it is possible to construct
a deterministic theory that at an operational level reproduces quantum
predictions then quantum indeterminism, and hence quantum randomness,
is not fundamental any more. Thus the question of existence of such
a theory is of considerable importance. 

The ontological models developed by Harrigan and Spekkens \cite{Harrigan'2010}
provide the framework to address the above question in a meaningful
way. Our work was motivated by a recent no-go result by Maroney \cite{Maroney'2012(1)},
and its subsequent analysis by Ballentine \cite{Ballentine'2014}.
Maroney proved, without additional assumptions, that it's impossible
to construct a maximally $\psi$-epistemic theory in dimensions greater
than two; i.e., there cannot be an ontological model, where the quantum
overlap between two state vectors can be completely accounted for
by the overlap of their respective probability distributions in the
space of ontic states. 

Ballentine showed that Maroney's result rules out ontological models
that are both reciprocal and deterministic. In other words, a reciprocal
ontological model is necessarily indeterministic. In this paper, we
proved that indeterminism in a reciprocal ontological model must be
of the order of quantum theory. Therefore, if we want to hold on to
objective reality then we should adopt an interpretation of wave function
close to $\psi$-ontic. \\
\\

\begin{acknowledgments}
SB, MB, SG, and GK would like to thank Ravi Kunjwal for many helpful
discussions. SB, MB, and GK would like to acknowledge The Institute
of Mathematical Sciences, Chennai for supporting a visit during which
part of this work was completed. AM thanks Council of Scientific and Industrial
Research, India for financial support through Senior Research
Fellowship (Grant No. 09/093(0148)/2012-EMR-I). SB is supported in part by DST-SERB
project SR/S2/LOP-18/2012. \end{acknowledgments}


\begin{thebibliography}{10}
\bibitem{Born'1926} M. Born, ``Quantenmechanik der Stoßvorgänge\textquotedbl{},
Z. Phys. \foreignlanguage{english}{\textbf{38}, 803 (1926).}

\bibitem{Neumann'1955} J. Von-Neumann, ``Mathematical Foundations
of Quantum Mechanics (English translation)'', \textit{Princeton University
Press}, 1955.

\bibitem{Jennewein'2000} T. Jennewein, U. Achleitner, G. Weihs, H.
Weinfurter, and A. Zeilinger, ``A fast and compact quantum random
number generator'', \href{http://dx.doi.org/10.1063/1.1150518}{Rev. Sci. Instrum.  {\bf 71}, 1675 (2000)} 

\bibitem{Stefanov'2000} A. Stefanov, N. Gisin, O. Guinnard, L. Guinnard,
and H. Zbinden, ``Optical quantum random number generator'', \href{http://www.tandfonline.com/doi/abs/10.1080/09500340008233380\#.UrPIldKhsnM}{J. Mod. Opt. {\bf 47}, 595 (2000)}.

\bibitem{Atsushi'2008} U. Atsushi \emph{et al.}, ``Fast physical
random bit generation with chaotic semiconductor lasers'', \href{http://www.nature.com/nphoton/journal/v2/n12/full/nphoton.2008.227.html}{Nature Photon. {\bf 2},  728 (2008)}.

\bibitem{Chaitin'1977} GL. Chaitin, ``Algorithmic Information Theory\textquotedbl{},
\href{http://domino.research.ibm.com/tchjr/journalindex.nsf/a3807c5b4823c53f85256561006324be/516a98adc4b295a785256bfa0067f802!OpenDocument}{IBM Journal of Research and Development {\bf 21}, 350 (1977)}.

\bibitem{Knuth'1981} D. Knuth, ``The Art of Computer Programming
Vol. 2, Semi-numerical Algorithms'', \textit{Addison-Wesley Publishing
Company}, 1981.

\bibitem{Butterfield'1998} J. Butterfield, ``Determinism and indeterminism'',
\href{https://www.rep.routledge.com/articles/determinism-and-indeterminism}{Routledge Encyclopedia of Philosophy {\bf 3}, 33 (1998)}.

\bibitem{Bell'1966} J. S. Bell, ``On the Problem of Hidden Variables
in Quantum Mechanics\textquotedbl{}, \href{http://journals.aps.org/rmp/abstract/10.1103/RevModPhys.38.447}{Rev. Mod. Phys. {\bf 38}, 447 (1966)}.

\bibitem{Mermin'1993} N. D. Mermin, ``Hidden variables and the two
theorems of John Bell\textquotedbl{}, \href{http://journals.aps.org/rmp/abstract/10.1103/RevModPhys.65.803}{Rev. Mod. Phys. {\bf 65}, 803 (1993)}.

\bibitem{KS'1967} S. Kochen and E. Specker, ``The Problem of Hidden
Variables in Quantum Mechanics\textquotedbl{}, \href{http://www.iumj.indiana.edu/IUMJ/fulltext.php?year=1968&artid=17004&volume=17}{Journal of Mathematics and Mechanics {\bf 17}, 59 (1967)}.

\bibitem{Harrigan'2010} N. Harrigan, R.W. Spekkens, ``Einstein,
Incompleteness, and the Epistemic View of Quantum States\textquotedbl{},
\href{http://link.springer.com/article/10.1007\%2Fs10701-009-9347-0}{Found. Phys. {\bf 40}, 125 (2010)}.

\bibitem{EPR'1935} A. Einstein, B. Podolsky, N. Rosen, ``Can Quantum-Mechanical
Description of Physical Reality Be Considered Complete?\textquotedbl{},
\href{http://link.aps.org/doi/10.1103/PhysRev.47.777}{Phys. Rev. {\bf 47}, 777 (1935)}.

\bibitem{Bohr'1935} N. Bohr, ``Can Quantum-Mechanical Description
of Physical Reality be Considered Complete?\textquotedbl{}, \href{http://link.aps.org/doi/10.1103/PhysRev.48.696}{Phys. Rev. {\bf 48}, 696 (1935)}.

\bibitem{Bohm'1952} D. Bohm, ``A suggested interpretation of quantum
theory in terms of `Hidden' variables. I\textquotedbl{}, \href{http://journals.aps.org/pr/abstract/10.1103/PhysRev.85.166}{Phys. Rev. {\bf 85}, 166 (1952)}.

\bibitem{Popper'1967} K.R. Popper, ``in Quantum Theory and Reality\textquotedbl{},
edited by M. Bunge (Springer, 1967).

\bibitem{Ballentine'1970} L.E. Ballentine, ``The Statistical Interpretation
of Quantum Mechanics\textquotedbl{}, \href{http://link.aps.org/doi/10.1103/RevModPhys.42.358}{Rev. Mod. Phys. {\bf 42}, 358 (1970)}.

\bibitem{Peres'1984} A. Peres, ``What is a state vector?\textquotedbl{},
\href{http://ajp.aapt.org/resource/1/ajpias/v52/i7/p644_s1?isAuthorized=no}{Am. J. Phys. {\bf 52}, 644 (1984)}.

\bibitem{Bohm'1993} D. Bohm and B. J. Hiley, The Undivided Universe
(Routledge, 1993).

\bibitem{Holland'1993} P. R. Holland, The Quantum Theory of Motion
(Cambridge, 1993).

\bibitem{Saunders'2010} S. Saunders, J. Barrett, A. Kent, and D.
Wallace, eds., Many Worlds? Everett, quantum theory and reality. (OUP,
2010).

\bibitem{Bassi'2003} A. Bassi and G. C. Ghirardi, ``Dynamical reduction models\textquotedbl{}, 
\href{http://www.sciencedirect.com/science/article/pii/S0370157303001030}{Physics Reports {\bf 379}, 257 (2003)}.

\bibitem{Maroney'2012(1)} O. J. E. Maroney, ``How statistical are
quantum states?\textquotedbl{}, \href{http://arxiv.org/abs/1207.6906}{arXiv:1207.6906 (2012)}.

\bibitem{Maroney'2012(2)} O. J. E. Maroney, ``A brief note on epistemic
interpretations and the Kochen-Speker theorem\textquotedbl{}, \href{http://arxiv.org/abs/1207.7192}{arXiv:1207.7192 (2012)}.

\bibitem{Lewis'2012} P. G. Lewis, D. Jennings, J. Barrett, and T.
Rudolph, ``Distinct Quantum States Can Be Compatible with a Single
State of Reality\textquotedbl{}, \href{http://journals.aps.org/prl/abstract/10.1103/PhysRevLett.109.150404}{Phys. Rev. Lett. {\bf 109}, 150404 (2012)}.

\bibitem{Aaronson'2013} S. Aaronson, A. Bouland, L. Chua, and G.
Lowther,``$\psi$-epistemic theories: The role of symmetry\textquotedbl{},
\href{http://journals.aps.org/pra/abstract/10.1103/PhysRevA.88.032111}{Phys. Rev. A {\bf 88}, 032111 (2013)}.

\bibitem{PBR'2012} M.F. Pusey, J. Barrett, T. Rudolph, ``On the
reality of the quantum state\textquotedbl{}, \href{http://www.nature.com/nphys/journal/v8/n6/full/nphys2309.html}{Nature Phys. {\bf 8}, 476 (2012)}.

\bibitem{Colbeck'2012} R. Colbeck, R. Renner, ``Is a System’s Wave
Function in One-to-One Correspondence with Its Elements of Reality?\textquotedbl{},
\href{http://journals.aps.org/prl/abstract/10.1103/PhysRevLett.108.150402}{Phys. Rev. Lett. {\bf 108}, 150402 (20102)}.

\bibitem{Ghirardi'2013} G-C Ghirardi and R. Romano, ``Ontological
Models Predictively Inequivalent to Quantum Theory\textquotedbl{},
\href{http://journals.aps.org/prl/abstract/10.1103/PhysRevLett.110.170404}{Phys. Rev. Lett. {\bf 110}, 170404 (2013)}.

\bibitem{Ballentine'2014} L. Ballentine, ``Ontological Models in
Quantum Mechanics: What do they tell us?\textquotedbl{}, \href{http://arxiv.org/abs/1402.5689}{arXiv:1402.5689 (2014)}.

\bibitem{Spekkens'2005} R.W. Spekkens, ``Contextuality for preparations,
transformations, and unsharp measurements\textquotedbl{}, \href{http://journals.aps.org/pra/abstract/10.1103/PhysRevA.71.052108}{Phys. Rev. A {\bf 71}, 052108 (2005)}.

\bibitem{Acin'2012} A. Acín, S. Massar, and S. Pironio, ``Randomness
versus Nonlocality and Entanglement\textquotedbl{}, \href{http://journals.aps.org/prl/abstract/10.1103/PhysRevLett.108.100402}{Phys. Rev. Lett. {\bf 108}, 100402 (2012)}.

\bibitem{Koenig'2009} R. Koenig, R. Renner, and C. Schaffner, ``Sampling
of Min-Entropy Relative to Quantum Knowledge\textquotedbl{}, \href{http://ieeexplore.ieee.org/xpl/articleDetails.jsp?tp=&arnumber=5895072&searchWithin\%3Dp_First_Names\%3AR\%26searchWithin\%3Dp_Last_Names\%3ARenner\%26matchBoolean\%3Dtrue\%26queryText\%3D\%28p_Authors\%3ARenner\%2C+R\%29}{IEEE Trans. Inf. Theory {\bf 55}, 4337 (2009)}.

\end{thebibliography}
\end{document}